\documentclass[showpacs,preprintnumbers,amsmath,amssymb]{revtex4}
\usepackage{graphicx}
\begin{document}

\title{    Quasiparticles for quantum dot array  in graphene and the associated Magnetoplasmons}
\author{Oleg L. Berman$^1$,  Godfrey Gumbs$^{2,3}$, and P.M. Echenique$^3$}
\affiliation{\mbox{$^1$Physics Department, New York City College of Technology of the
City University of New York} \\ 300 Jay Street, Brooklyn, NY 11201 \\
\mbox{$^2$ Department of Physics and Astronomy,
Hunter College of the City University of New York} \\
695 Park Avenue, New York, NY 10065\\
\mbox{$^3$ Departamento de fisica de materials and
centro mixto CSIC-UPV }
 \\   Donostia International Physics Center (DIPC)\\ P. de Manuel Lardizabal, 4, 20018 San Sebastián,
Basque Country, Spain}

\date{\today}

\begin{abstract}
We  calculate the  low-frequency magnetoplasmon excitation spectrum   for a square
array of  quantum dots on a two-dimensional (2D) graphene layer.
 The confining potential is linear in the distance from the center of the quantum dot.
 The electron eigenstates  in a magnetic  field  and  confining  potential  are
mapped onto a  2D plane  of electron-hole pairs in an effective magnetic field
without any confinement. The tight-binding model for the array of quantum dots
leads to a wavefunction with  inter-dot mixing of the quantum numbers associated with an
isolated quantum dot.  For chosen  confinement,
magnetic field, wave vector and frequency, we plot the dispersion
equation as a function of the period $d$ of the lattice. We obtain
those values of  $d$ which yield collective plasma excitations. For
the allowed transitions between the valence and conduction bands in
our calculations, we obtain plasmons when $d \lesssim 100 {\AA}$.

\end{abstract}


\maketitle

\section{Introduction}
\label{intro}

A two-dimensional (2D) honeycomb lattice of carbon atoms that form the basic planar
structure in graphite (graphene) has recently been produced\cite{Novoselov1,Zhang1}.
  Unusual many-body effects  in graphene have been attributed to    low-lying excitations
in the vicinity of the Fermi level\cite{blinowski,Shung1,Shung2,DasSarma}.  The influence of an
   external magnetic field on the many-electron properties    of graphene results in
an unusual quantum Hall effect. The integer quantum Hall effect   (IQHE) was
discovered in graphene    in recent experiments\cite{Novoselov2,Zhang2,Zhang3}.
The quantum Hall ferromagnetism in graphene has been studied theoretically \cite{Nomura}. In addition, the influence of magnetic field on the Bose-Einstein
condensation and superfluidity of indirect magnetoexcitons in graphene bilayer has been studied \cite{BLG}.
 The spectrum of plasmon excitations in a single graphene layer immersed
in a material with effective dielectric   constant $\varepsilon_b$ in the absence of
    magnetic field  ($B=0$) was calculated in   Refs.\ \cite{Shung1,Shung2,Hwang}.
(See also \cite{new1,new2,new3,new4}.)  The collective
plasma excitations   in  layered graphene  structures in high magnetic field were
       obtained  in Ref.\ \cite{berman_gumbs_lozovik} where the instability of these
modes was investigated. Recently, several works were reported for
quantum dots in graphene \cite{dots4,dots5,dots6}. The transport
characteristics of quantum dot devices etched entirely in graphene
have been studied experimentally \cite{Novoselov_dots}. According to
Ref.\ \cite{Novoselov_dots}, these quantum dots at large sizes
behave as conventional single-electron transistors. This is just one
of the areas of research in the fast-growing field of the electronic
transport , thermal and optical properties of graphene which may
have important device applications because of its high mobility
\cite{guinea}.
 The single-particle states and collective modes in semiconductor quantum dots have
long been a subject of interest to both theoreticians and
experimentalists \cite{dots1,dots3,Maksym}. The spectrum of collective plasma modes
 in a 2D array of quantum dots in semiconductors has been calculated \cite{Que}.

In this paper, we obtain  the dispersion relation for
magnetoplasmons in  a square array of  quantum dots formed by a
confining potential  which is a linear function of distance from the
center of the quantum dot.  A  perpendicular magnetic field
$\mathbf{B}$ is applied. The conditions for the existence of the
collective excitations will be  investigated.

This paper is organized as follows. In Sec.\ \ref{single}, we
introduce the quasiparticle electron-hole representation for the
eigenvalue problem for an electron in a single graphene quantum dot
in magnetic field. In Sec.\ \ref{array_dot} we represent the
calculations of the magnetoplasmon spectrum in a quantum dot array
in graphene. A brief discussion of the results of our calculations
plasmon instabilities in graphene is presented in Sec.\ \ref{disc}.

\section{Single-electron States for a  graphene quantum dot in magnetic field}
\label{single}

We now consider an electron  in a graphene layer in the presence of
a perpendicular magnetic field $\mathbf{B}$ and a confining
potential   $U(r^\prime)= \lambda_{0}|r^\prime|$, where
$\lambda_0>0$ is the slope of the potential. The effective-mass
Hamiltonian of an electron in the absence of scatterers  in one
valley in graphene  located in the $xy$-plane is given
by a $2\times 2$ matrix Hamiltonian $\hat{H}_{(0)}$ with zero along
the diagonal and off-diagonal elements $\hat{\pi}_{x} \mp  i \hat{\pi}_{y}$ \cite{Ando,Jain}.
Here, we neglect the Zeeman splitting and assume energy degeneracy
with respect to the two valleys. Also, in our notation,  $\hat{\mathbf{\pi}} = -i\hbar\nabla +
e\mathbf{A}$,  $\mathbf{A}$ is the vector potential of an electron,
$v_{F} = \sqrt{3}at/(2\hbar)$ is the Fermi velocity of electrons
with $a =2.566 \AA$  denoting the lattice constant and $t \approx
2.71 eV$ is the overlap integral between  nearest-neighbor carbon atoms in
graphene \cite{Lukose}. The eigenvalue problem of an electron in a
linear confining potential can be mapped onto one for a
noninteracting electron-hole pair in an effective magnetic field
$\tilde{B}_{\rm eff}$ under conditions we introduce below.

The eigenfunction $\psi_{\tau}$ for the Hamiltonian $\hat{H}_{(0)}$
 for an  electron-hole pair  in an effective magnetic
field $\tilde{B}_{\rm eff}$, is also the eigenfunction of the
magnetic momentum $\hat{\mathbf{P}}$  has the
form \cite{Gorkov,Lerner,Kallin} and is given by

\begin{eqnarray}\label{psi1}
\psi_{\mathbf{P}}(\mathbf{R},\mathbf{r}) = \exp\left[
\frac{i}{\hbar}\mathbf{R}\cdot \left(\mathbf{P} + e
[\tilde{\mathbf{B}}_{\rm eff}\times
\mathbf{r}]\right)\right]\tilde{\Phi}(\mathbf{r} -
\mathbf{\rho}_{0})\ ,
\end{eqnarray}
where $\mathbf{R}= (\mathbf{r}_{e} + \mathbf{r}_{h})/2$, $\mathbf{r}
= \mathbf{r}_{e} - \mathbf{r}_{h}$ and $\mathbf{\rho}_{0}
=[\tilde{\mathbf{B}}_{\rm eff}\times\mathbf{P}]/(e \tilde{B}_{\rm eff}^{2})$.
The cylindrical gauge for vector potential is used with
$\mathbf{A}_{e(h)} = 1/2  [\tilde{\mathbf{B}}_{\rm eff}\times
\mathbf{r}_{e(h)}]$. The wavefunction of the relative coordinate
$\tilde{\Phi}(\mathbf{r})$ can be expressed in terms of the 2D
harmonic oscillator eigenfunctions $\Phi_{n_{1},n_{2}}(\mathbf{r})$.
For an electron in Landau level $n_{+}$ and a hole in level $n_{-}$,
the four-component wavefunctions for the relative motion
in graphene are \cite{Fertig}

\begin{eqnarray}\label{electron}
&& \tilde{\Phi}_{n_{+},n_{-}}(\mathbf{r}) =   \left(
\sqrt{2}\right)^{\delta_{n_{+},0}+\delta_{n_{-},0}-2} \nonumber
\\
&& \times \left(
\begin{array}{c} s_{+}s_{-}
\Phi_{|n_{+}|-1,|n_{-}|-1}(\mathbf{r})\\
s_{+}\Phi_{|n_{+}|-1,|n_{-}|}(\mathbf{r})\\
s_{-}\Phi_{|n_{+}|,|n_{-}|-1}(\mathbf{r})\\
\Phi_{|n_{+}|,|n_{-}|}(\mathbf{r})
\end{array}\right)\ ,
\end{eqnarray}
where $s_{\pm} = \mathrm{sgn} (n_{\pm})$. The corresponding energy
of the electron-hole pair $E_{n_{+},n_{-}}$  of the Hamiltonian
$\hat{H}_{(0)}$  is given by \cite{Fertig}
\begin{eqnarray}
\label{spgr} && E_{n_{+},n_{-}} = \sqrt{2}\ \left( (\hbar
v_F/r_{B}^{\ast}) \right) \nonumber \\
&& \times \left[\mathrm{sgn}(n_{+})\sqrt{|n_{+}|} -
\mathrm{sgn}(n_{-})\sqrt{|n_{-}|}\right]\ ,
\end{eqnarray}
 where $r_{B}^{\ast}=
\sqrt{\hbar/(e\tilde{B}_{\rm eff})}$ is an effective magnetic
length. The 2D harmonic oscillator wavefunctions
$\Phi_{n_{1},n_{2}}(\mathbf{r})$ are given by \cite{Fertig}
\begin{eqnarray}\label{electron_d}
&& \Phi_{n_{1},n_{2}}(\mathbf{r}) =
(2\pi)^{-1/2}2^{-|m|/2}\frac{\tilde{n}!}{\sqrt{n_{1}!n_{2}!}}
\frac{1}{r_{B}^{*}} \mathrm{sgn}(m)^{m} \nonumber
\\ && \times \frac{r^{|m|}}{r_{B*}^{|m|}} \exp\left[-im\phi -
\frac{r^{2}}{4r_{B}^{*2}}\right]
L_{\tilde{n}}^{|m|}\left(\frac{r^{2}}{2r_{B}^{*2}}\right) \ ,
\end{eqnarray}
where $L_{\tilde{n}}^{|m|}(x)$  is a Laguerre polynomial, $m = n_{1}
-n_{2}$, $\tilde{n} = \min(n_{1},n_{2})$ and $\mathrm{sgn}(m)^{m}\to
1$ for $m=0$.  The electron-hole eigenfunction given by Eqs.\
(\ref{psi1}), (\ref{electron}),  (\ref{electron_d}) along with the
energy eigenvalues  are the same as the
eigenfunction and the eigenvalue of the Hamiltonian  $\hat{H}_{(0)}$
 for an electron in a quantum dot in external magnetic
field. The magnetic momentum $\mathbf{P} = - e
[\tilde{\mathbf{B}}_{\rm eff}\times \mathbf{r}]$ must be the same
for the electron and the electron-hole pair at fixed relative
coordinate $\mathbf{r}$. The effective magnetic field
$\tilde{B}_{\rm eff}$ depends on the slope  $\lambda_{0}$ of the
confining potential as well as the external magnetic field  $B$ and
is given by
\begin{eqnarray}
\label{mag_eff}
 \tilde{B}_{\rm eff} = \left(B^{2} + (4
\lambda_0^2/(e^2v_F^2)\right)^{1/2} \ .
\end{eqnarray}
In the case when there is no electron confinement, i.e., $\lambda_0
= 0$,  the effective magnetic field is the same as the applied
magnetic field and $\tilde{B}_{\rm eff} = B$. If there is no
external magnetic field, i.e.,  $B = 0$, then there is still an
effective magnetic field due to   the electron confinement and is
given by $\tilde{B}_{\rm eff} = 2 \lambda_0/(ev_F)$. The coordinate
of the electron-hole relative motion $r =
|\mathbf{r}_{e}-\mathbf{r}_{h}|$ is related to the distance
$r^\prime$ of an electron from the center of the confining potential
by $r = r^\prime/2$.

It should be emphasized that the electrostatic potential for a
graphene quantum dot was chosen to have a linear dependence on the
coordinate variable $|r^\prime|$. This was done because  the
eigenvalue problem of the Hamiltonian $\hat{H}_{(0)}$ described
by the Dirac-like equation with a linear potential can be reduced
to the Klein-Gordon-type equation  with a parabolic potential
$U(r^\prime)= \lambda_{0}^{2}r^{\prime\ 2}$. This Klein-Gordon-type
equation can then be mapped onto the electron-hole problem in an
effective magnetic field $\tilde{B}_{\rm eff}$ defined above
 under the conditions presented.

\section{Magnetoplasmons for a quantum dot array in graphene}
\label{array_dot}

\begin{figure}
\includegraphics[width=2.2in]{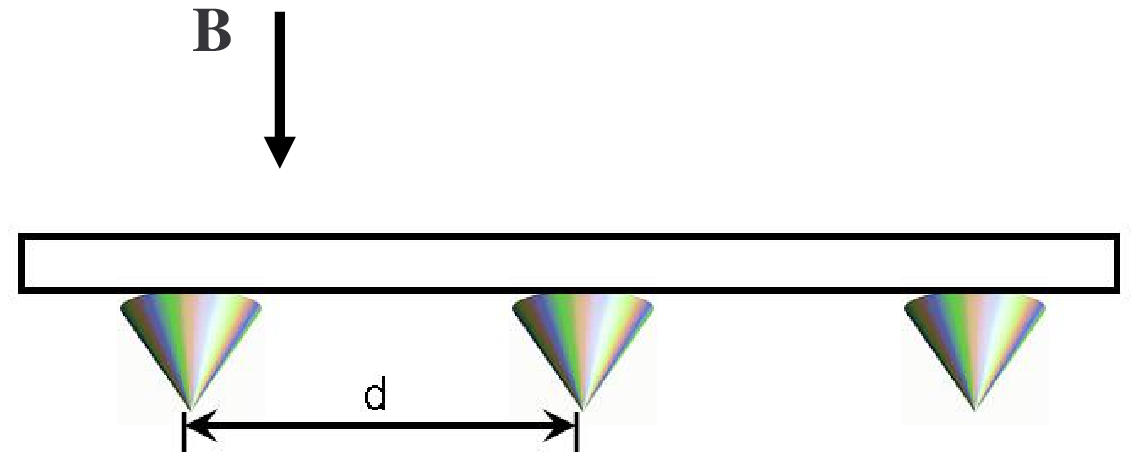}
\caption{A periodic 2D array of quantum dots in a graphene plane
formed by   linear confinement with a spacing $d$ in a
perpendicular magnetic field $\mathbf{B}$.} \label{array}
\end{figure}

We now turn our attention to an infinite periodic 2D array of
quantum dots in a graphene plane, shown schematically in Fig.\
\ref{array}.  The quantum dots are formed by a periodic linear
confining potential defined by  $U(r^\prime)= \sum_j
\lambda_{0}|\mathbf{r}^\prime- \mathbf{r}_{j}|$, where
$\mathbf{r}_{j}$ is the position vector of a quantum dot. We
consider this array with the period $d$ in a perpendicular magnetic
field $\mathbf{B}$ as shown schematically in  Fig.\ \ref{array}. For
this system, we apply the tight-binding approximation
\cite{Que,gumbs,Bottger} (see Eq.~(2.1) in Ref.\ \cite{Que})

\begin{eqnarray}\label{tba}
\Psi_{{\bf k}}(x,y)=\left|\mathbf{k},\alpha\right\rangle =
\sum_{\mathbf{r}_{j}}e^{i\mathbf{k}\cdot\mathbf{r}_{j}}\psi_{\alpha}(\mathbf{r}-
\mathbf{r}_{j})\exp\left(-\frac{ie}{\hbar}\mathbf{A}\cdot\mathbf{r}_{j}\right)\ ,
\end{eqnarray}
where $\mathbf{k}$ is an in-plane wave vector,  $\psi_{\alpha}$ is the
electron eigenfunction in one   quantum dot (which can be
represented by the wavefunction of an electron-hole pair in an
effective magnetic field as $\psi_{\alpha} =
\tilde{\Phi}_{n_{+},n_{-}}(\mathbf{r}/2)$, where
$\tilde{\Phi}_{n_{+},n_{-}}(\mathbf{r}/2)$ is given by Eq.\
(\ref{electron}). The index  $\alpha = \{n_{+},n_{-}\}$ is a
composite quantum number with $n_{+}$ and $n_{-}$ labeling the
electron and hole energy levels, respectively. Also, $\mathbf{A}$
is the vector potential for the
externally applied magnetic field $\mathbf{B}$.

Note that  the electron wave function in periodic systems in
a magnetic field is the eigenfunction of both the Hamiltonian and
the operator of the magnetic translation. The electronic properties
in the presence of magnetic field are determined  by the magnetic
flux through a unit cell. \cite{Thouless} If this flux (measured in
flux quantum $\phi_0=h/e$) is a rational number $\Phi/\phi_0 = p/q
=Ba^2e/2\pi\hbar $ where $p$ and $q$ are prime integers, then the
electron wave function which is also  an eigenfunction of the
magnetic translation operator \cite{LP} obeys the Bloch-Peierls
conditions in the Landau gauge  which may be expressed as\cite{Thouless}

\begin{equation}
\Psi_{{\bf k}}(x,y)=\Psi_{{\bf
k}}(x+qa,y+a)e^{-ik_xqa}e^{-ik_ya}e^{-2\pi i py/a} \ . \label{gg1}
\end{equation}
For integer $m_1,m_2$, the vectors ${\bf a}_m(m_1qa,n_2a)$ define
the magnetic lattice of the crystal. The magnetic Brillouin zone is
defined by the inequalities $-\pi/qa\leq k_x\leq \pi/qa$ and
$\pi/a\leq k_y\leq k_y\leq \pi/a$.

Let us demonstrate that the  electron wavefunction given
by Eq.\ (\ref{tba}) obeys the Bloch-Peierls conditions in the Landau
gauge. In this  gauge,  the vector potential may be chosen as
${\bf A}=(0,Bx,0)$. Then, Eq.\ (\ref{tba}) yields

\begin{eqnarray}
\Psi_{{\bf k}}(x+qa,y+a)=
\sum_{\mathbf{r}_{j}}e^{i\mathbf{k}\cdot\mathbf{r}_{j}}\psi_{\alpha}((x+qa)\hat{i}+(y+a)\hat{j}-
\mathbf{r}_{j})\exp\left(-\frac{ie}{\hbar}B(x+qa)\hat{j}\cdot\mathbf{r}_{j}\right)\
. \label{gg2}
\end{eqnarray}
Now, let ${\bf r}_j-qa\hat{i}-a\hat{j}={\bf r}_j^\prime$. Then, Eq.\
(\ref{gg2}) becomes
\begin{eqnarray}
\Psi_{{\bf k}}(x+qa,y+a)&=&
\sum_{\mathbf{r}_{j}^\prime}e^{i\mathbf{k}\cdot(\mathbf{r}_{j}^\prime+qa\hat{i}+a\hat{j})}
\psi_{\alpha}({\bf r}-
\mathbf{r}_{j}^\prime)\exp\left(-\frac{ie}{\hbar}B(x+qa)\hat{j}\cdot(\mathbf{r}_{j}^\prime+qa\hat{i}+a\hat{j})\right)
\nonumber\\
&=& e^{ik_xqa}e^{ik_ya}
\sum_{\mathbf{r}_{j}^\prime}e^{i\mathbf{k}\cdot\mathbf{r}_{j}^\prime}\psi_{\alpha}(x\hat{i}+y\hat{j}-
\mathbf{r}_{j}^\prime)
\nonumber\\
&\times& \exp\left( \frac{-ie}{\hbar}{\bf A}\cdot{\bf r}_j^\prime
-2\pi ip- 2\pi i \frac{p}{q}\frac{x}{a}-2\pi i p\left( \frac{{\bf
r}_j\cdot\hat{j}}{a} \right)\right)
\nonumber\\
&=& e^{ik_xqa}e^{ik_ya}\exp\left( -2\pi i
\frac{p}{q}\frac{x}{a}\right) \Psi_{{\bf k}}(x,y) \ . \label{gg3}
\end{eqnarray}
But, this result in  Eq.\ (\ref{gg3}) is equivalent to  Eq.\
(\ref{gg1}). The reason is that they coincide when we rotate axes in
the frame before doing the displacement, i.e., $x$-axis $\to
-y$-axis or, replacing $x\to -y/q$  on the right-hand-side of Eq.\
(\ref{gg3}). This is now explicitly demonstrated below. Using the
gauge ${\bf A}=(-By,0,0)$, we obtain
\begin{eqnarray}
\Psi_{{\bf k}}(x+qa,y+a)&=&
\sum_{\mathbf{r}_{j}^\prime}e^{i\mathbf{k}\cdot(\mathbf{r}_{j}^\prime+qa\hat{i}+a\hat{j})}
\psi_{\alpha}({\bf r}-
\mathbf{r}_{j}^\prime)\exp\left(-\frac{ie}{\hbar}B(-y-a)\hat{i}\cdot(\mathbf{r}_{j}^\prime+qa\hat{i}+a\hat{j})\right)
\nonumber\\
&=& e^{ik_xqa}e^{ik_ya}
\sum_{\mathbf{r}_{j}^\prime}e^{i\mathbf{k}\cdot\mathbf{r}_{j}^\prime}\psi_{\alpha}(x\hat{i}+y\hat{j}-
\mathbf{r}_{j}^\prime)
\nonumber\\
&\times& \exp\left( \frac{-ie}{\hbar}{\bf A}\cdot{\bf
r}_j^\prime+2\pi ip+ 2\pi i p \frac{y}{a}+2\pi i \frac{p}{q}\left(
\frac{{\bf r}_j\cdot\hat{i}}{a} \right)\right)
\nonumber\\
&=& e^{ik_xqa}e^{ik_ya}\exp\left( 2\pi i p\frac{y}{a}\right)
 \nonumber\\
 &\times&\sum_{\mathbf{r}_{j}^\prime}e^{i\mathbf{k}\cdot\mathbf{r}_{j}^\prime}\psi_{\alpha}(x\hat{i}+y\hat{j}-
\mathbf{r}_{j}^\prime) \exp\left\{ 2\pi i \frac{p}{q}\left(
\frac{{\bf r}_j^\prime \cdot\hat{i}}{a} \right)\right\} \
.\label{gg4}
\end{eqnarray}
The last line in Eq.~(\ref{gg4}) satisfies the Eq.\ (\ref{gg1})
perfectly if we take ${\bf k}$ to be ${\bf k}$ modulo $2\pi/aq$.
Therefore, we conclude that the electron wavefunction given by Eq.\
\ref{tba} obeys the Bloch-Peierls conditions in the Landau gauge. In
Ref.\ [\onlinecite{Perov}] quantum states of a similar system were
considered. The energy spectrum in this case consists of magnetic
subbands defined in a magnetic Brillouin zone. The structure of
magnetic subband depends on the number of magnetic flux penetrating
the unit cell. The  wave function given by Eq.~(\ref{tba})
corresponds to one in Ref.\ [\onlinecite{Perov}] if we do not
include spin-orbit coupling in our model.

We do not consider magnetic fields where the flux is $p/q$ in our numerical calculation, since this would involve a formalism like the one presented in Refs.~[\onlinecite{gumbs2,gumbs3}]. Furthermore, the resulting fractal nature of the energy spectrum under these conditions can only be realized numerically for magnetic fields which are several tens of Tesla. This is beyond the interest and scope of the present paper.

Following the procedure adopted in  Ref.\ \cite{Que}, we introduce
the quantity

\begin{eqnarray}
\label{pol} \Pi_{\alpha \beta}(\omega) &=&
\left(\frac{g_{s}g_{v}}{2\pi r_{B}^{\ast\ 2}} \right) \frac{f_0(E_\alpha)
- f_0(E_\beta)}{\hbar\omega+E_{\alpha} - E_{\beta} +i0^+ }\  ,
\end{eqnarray}
where $g_{s}=2$ and $g_{v}=2$ are the spin and  valley degeneracies
in graphene, $\omega$ denotes frequency,  $\alpha = (n_{+},n_{-})$,
$\beta = (n_{+}',n_{-}')$, $ E_{\alpha (\beta)}$ are the
electron-hole pair eigenenergies  for
the Landau levels in an effective magnetic field. Also, $f_0(E)$ is the occupation function of the
two-particle state.
 At high temperatures or  weak magnetic field, when the separation between Landau
levels is small so that $k_BT \gg \hbar v_F /r_B^{\ast} $, the
occupation of the Landau levels is given by the Fermi-Dirac
distribution function $f_0(E_\alpha)=
\left(\exp\left[E_{\alpha}/(k_B T)\right] + 1 \right)^{-1}$.

We now introduce the overlap integral $F_{\alpha \beta}
(\mathbf{q})$  involving the eigenstates with labels $\alpha =
\{n_{+},n_{-}\}$ and $\beta = \{n_{+}^\prime,n_{-}^\prime\}$ through
the equation \cite{Que}

\begin{eqnarray}
\label{over} && F_{\alpha \beta} (\mathbf{q}) = \int
d^{2}\mathbf{r}\ e^{i\mathbf{q}\cdot
\mathbf{r}}\psi_{\alpha}^{\ast}(\mathbf{r})\psi_{\beta}(\mathbf{r})
= s_{+}s_{+}^\prime s_{-}s_{-}^\prime \ \nonumber \\
&& \times  F_{|n_{+}|-1,|n_{-}|-1,|n_{+}^\prime|-1,|n_{-}^\prime
|-1}^{(0)} (\mathbf{q}) \nonumber \\ && + s_{+}s_{+}^\prime\
F_{|n_{+}|-1,|n_{-}|,|n_{+}^\prime|-1,|n_{-}^\prime|}^{(0)}
(\mathbf{q}) \nonumber \\ && + s_{-}s_{-}^\prime\
F_{|n_{+}|,|n_{-}|-1,|n_{+}^\prime|,|n_{-}^\prime|-1}^{(0)}
(\mathbf{q}) \nonumber \\ && +
F_{|n_{+}|,|n_{-}|,|n_{+}'|,|n_{-}^\prime|}^{(0)}(\mathbf{q})\ ,
\end{eqnarray}
where
\begin{eqnarray}
\label{over1}
F_{n_{1},n_{2},n_{1}^\prime,n_{2}^\prime}^{(0)}(\mathbf{q}) = \int
d^{2}\mathbf{r}\
e^{i\mathbf{q}\cdot\mathbf{r}}\Phi_{n_{1}',n_{2}'}^{\ast}(\mathbf{r})\Phi_{n_{1},n_{2}}(\mathbf{r})\label{formfac}
\end{eqnarray}
and $\Phi_{n_{1},n_{2}}(\mathbf{r})$ is given by Eq.\
(\ref{electron_d}). The angular integral in Eq.\ (\ref{formfac}) can
be carried out using the result
\begin{eqnarray}
\label{ang}
\int_0^{2\pi} d\theta\
e^{i\beta\cos(\theta)-im\theta}=2\pi(-i)^{m}J_{m}(-\beta)
\end{eqnarray}
with  $m$  an integer. If $m$ is even, $J_{m}(-\beta)=J_{m}(\beta)$.

We employ the procedure introduced in  Ref. \cite{Que} for
determining  the dispersion relation $\omega = \omega (\mathbf{q})$
for the magnetoplasmon excitation frequencies.   This can be
calculated from the condition of the vanishing of the determinant of
the matrix with elements $C_{ij}(\mathbf{q},\omega)$. In this
notation, the subscript $i$ denotes the pair of quantum numbers
$\alpha$ and  $\beta$ for two different composite electron-hole
pairs. Also,  $j$  denotes the quantum numbers $\alpha^\prime$ and
$\beta^\prime$ for two other composite electron-hole pairs.  The
matrix elements $C_{ij}$ are defined as

\begin{eqnarray}
\label{Cij}
&& C_{ij}(\mathbf{q},\omega) =  \delta_{ij} \nonumber \\
&& - \Pi_{i} (\omega) \sum_{\mathbf{G}} \frac{2\pi
e^{2}}{\varepsilon_s |\mathbf{q} + \mathbf{G}|} F_{i}(\mathbf{q} +
\mathbf{G})F_{j}^{\ast}(\mathbf{q} + \mathbf{G}) \ .
\end{eqnarray}
In this notation, $\varepsilon_s=4\pi\epsilon_0\epsilon_b$, where
$\epsilon_b$ is the average dielectric constant of the medium where the
layer of graphene is embedded.  We also introduced
$\mathbf{G}=2\pi/d\left(n_x,n_y \right)$ which is a reciprocal
lattice vector of the square array of
 quantum dots with period $d$ and $n_x,\ n_y=0,\pm 1,\pm 2,\cdots$.

\begin{figure}
\includegraphics[width = 2.2in]{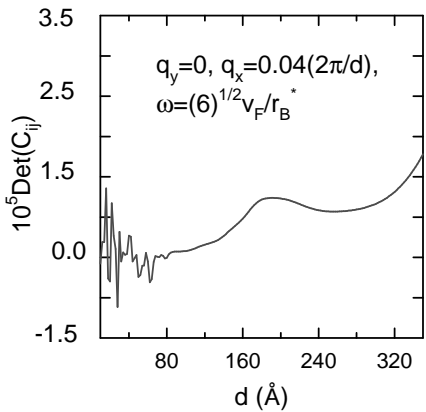}
\caption{$\det [ C_{ij}(\mathbf{q},\omega)]$ as a function of an
array spacing $d$ calculated using  Eq.\ (\ref{Cij}). We chose   the
slope of the confinement  $\lambda _{0} = 10^{-2} eV/{\AA} $,
magnetic field $B = 1 T$, $\epsilon_b=2.5$ and the temperature T=0
K.  The values of spacing $d$ resulting in $\det [
C_{ij}(\mathbf{q},\omega)] = 0$ correspond to the existence of
magnetoplasmon excitations at chosen frequency and wave vector.}
\label{plasmon}
\end{figure}

In solving the dispersion equation for plasmon excitations, we  chose
the slope of the  confining potential, magnetic field, wave vector, frequency,
temperature  and  background dielectric constant. We then varied the period
$d$ and calculated the determinant
$\det [ C_{ij}(\mathbf{q},\omega)] $ obtained from Eq.\ (\ref{Cij}).
The zeros of the determinant correspond to the collective magnetoplasmon
excitations. As is shown in Fig.\ \ref{plasmon},  allowed plasmon resonances
appear in the system at specific values of the dot spacing $d$ corresponding
to the condition $\det [ C_{ij}(\mathbf{q},\omega)] = 0$. The
results  of these  calculations were obtained by taking into account only
transitions between the Landau levels $n = -1, 0, 1$ .
The interdot separations where the plasmons can be excited are given by
 $d \lesssim 100 {\AA}$. The excitation spectrum of the collective
modes preserves the periodicity of the lattice, even with the mixing
of the quantum numbers.

\section{Discussion}

\label{disc}

\begin{figure}
\includegraphics[width = 2.2in]{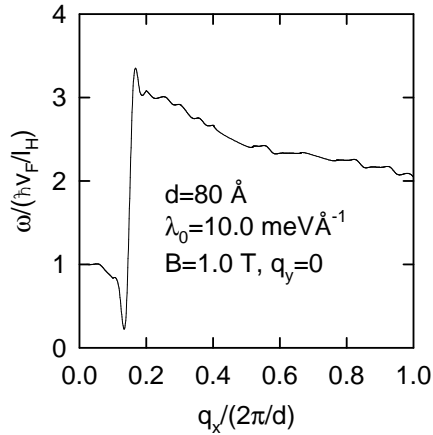}
\caption{We show in this figure the plasmon    excitation frequency
as a function of $q_x$ for $q_y=0$.  The slope of the confinement
$\lambda _{0} = 10^{-2} eV/{\AA} $, magnetic field $B = 1 T$,
$\epsilon_b=2.5$ and the temperature T=0 K.  } \label{fig3}
\end{figure}

In Fig.\ \ref{fig3}, we present the dispersion relation for plasmons
in a 2D array of quantum dots in a perpendicular magnetic field at zero
temperature. The parameters used in the calculation are given in the
figure.  The frequency is plotted as a function of $q_x$ with $q_y=0$.
As $q_x$ increases in the long wavelength limit, the group velocity is
initially small but then rapidly becomes negative for a range of
wave vector.  Then, over a narrow range of wave vector, the
frequency increases before it again decreases as it approaches the
Brillouin zone boundary. The finite frequency of excitation in the
long wavelength limit is due to the allowed transition between the
valence and conduction band. In  calculating the dispersion, we must
solve for the zeros of the $(2N+1)^4\times (2N+1)^4$-dimensional
$\det [ C_{ij}(\mathbf{q},\omega)]$ as a function of  frequency, where
$i,j=0,\pm 1,\pm 2,\cdots,\pm N$ are the energy level labels in the
valence and conduction bands. Thus, when we include the $0,\pm 1$
levels only, the determinant has dimension eighty-one.   We note that the
negative dispersion must be due to the effective magnetic field
since the group velocity is always positive for a graphene sheet in the
absence of magnetic field \cite{Shung1,DasSarma,new4}.
The commensurability effect between the electron cyclotron orbits and
the period of the quantum dot array can be explained by the oscillations in
the plasma dispersion shown in Fig.\ \ref{fig3}. The interaction term
decreases as the wave vector increases, thereby resulting in the
decrease in plasmon excitation energy as $q_x$ approaches the Brillouin
zone boundary. Since each value of inter-dot separation has its own set of
eigenstates, the dispersion relation could be drastically changed by
choosing a different value of $d$. We note that a zero-gap graphite
sheet only excites the inter-band electron-hole excitations at zero
temperature.  At finite temperature, there will be intra-band excitations
which induce intra-band plasmon modes.

In summary, the  methodological achievement of the approach
presented in this paper is the mapping  of  the single-electron
eigenfunctions (Eq.\ (\ref{electron})) and eigenenergies  in
magnetic field in linear confinement in graphene onto the
eigenfunctions and eigenenergies of a new set of quasiparticles of
quasielectrons and quasiholes, without  confinement in an effective
magnetic field depending on the slope of the confinement.  The
quasiparticle technique is also well suited for calculating the
different linear responses (e.g., conductivity, linear paramagnetic
susceptibility, optical response, etc.) which can be found by
computing the average $\langle G^R(\omega,{\bf r})G^A(\omega,-{\bf
r},)\rangle$ involving the retarded and advanced Green's functions.
Also, the calculation of the electronic heat capacity, which, as
usual, is found from the the temperature derivative of the mean
energy, utilizes the quasiparticle energies. Thus. the scope of
application of the quasiparticle concept is wide and can be employed
in both transport, thermodynamic and optical studies. We showed that
the dispersion equation yields allowed plasmon excitations for
specific inter-dot separation at chosen wave vector and frequency.
Our dispersion curve shows negative group velocity which is
generated by the effective magnetic field arising from the external
magnetic field and confining potential.

\acknowledgments This work is supported by contract FA
9453-07-C-0207 of AFRL.

\begin{thebibliography}{99}


\bibitem{Novoselov1} K. S. Novoselov {\it et al.}, Science {\bf 306}, 666 (2004).

\bibitem{Zhang1} Y. Zhang, J.~P. Small, M.~E.~S. Amori, and P. Kim, \prl {\bf 94}, 176803 (2005).

\bibitem{blinowski} J. Blinowski, N. H. Hau, C. Rigaux, J. P. Vieren, R. L. Toullee, G. Furdin, A. Herold and J. Melin, J. Phys. (Paris)  {\bf 41},  47 (1980).

\bibitem{Shung1} Kenneth W. -K. Shung, \prb {\bf  34}, 979  (1986).


\bibitem{Shung2} Kenneth W. -K. Shung, \prb {\bf 34}, 1264  (1986).


\bibitem{DasSarma} S. Das Sarma, E. H. Hwang, and W.- K. Tse, \prb {\bf  75}, 121406(R) (2007).

\bibitem{Novoselov2} K. S. Novoselov {\it et al.}, Nature (London) {\bf 438}, 197 (2005).

\bibitem{Zhang2} Y. B. Zhang {\it et al.}, Nature (London) {\bf 438}, 201 (2005).

\bibitem{Zhang3} Y. Zhang {\it et al.}, \prl {\bf 96}, 136806 (2006).

\bibitem{Nomura} K. Nomura and A. H. MacDonald, \prl  {\bf 96}, 256602 (2006).

\bibitem{BLG} O.~L. Berman, Yu.~E. Lozovik and G. Gumbs, \prb {\bf 77}, 155433 (2008).

\bibitem{Hwang} S. Das Sarma and  E. H. Hwang,  \prl {\bf 81}, 4216 (1998).

\bibitem{new1}     S. Gangadharaiah, A. M. Farid, and E. G. Mishchenko,
\prl  {\bf 100}, 166802 (2008).

\bibitem{new2} B. Wunsch, T. Stauber, F. Sols, F. Guinea,
New Journal of Physics {\bf 8}, 318 (2006).

\bibitem{new3} M. F. Lin, C. S. Huang, and D. S. Chuu,  \prb  {\bf 55}, 13961 (1997).

\bibitem{new4} Ming-Fa Lin and  Feng-Lin Shyu,  Jour. Phys. Soc. Japan {\bf 69},  607 (2000).


\bibitem{berman_gumbs_lozovik}  O. L. Berman, G.  Gumbs and Yu. E.
Lozovik, \prb {\bf 78}, 085401 (2008).


\bibitem{dots4} Milton Pereira, P. Vasilopoulos, and F. M. Peeters, Nano Lett. {\bf 7}, 946 (2007).

\bibitem{dots5} N. M. R. Peres, A. H. Castro Neto, F. Guinea, \prb {\bf  73},   241403(R) (2006).

\bibitem{dots6} P. G. Silvestrov and  K. B. Efetov,  \prl {\bf 98}, 016802 (2007).

\bibitem{Novoselov_dots}  L. A. Ponomarenko, F. Schedin, M. I. Katsnelson, R. Yang, E. H. Hill, K. S. Novoselov, and A.
K. Geim, Science {\bf 320}, 356 (2008).


\bibitem{guinea} A. H. Castro Neto,  F. Guinea, N. M. R. Peres,  K. S. Novoselov, and A. K. Geim,
Arxiv; 0709.1163v1 (2007).



\bibitem{dots1} L. Brey, N. F.  Johnson, and B. I. Halperin, \prb {\bf  40}, 10647 (1989).

\bibitem{dots3} F. M. Peeters, \prb {\bf  42}, 1486 (1990).

\bibitem{Maksym} P. A. Maksym and T. Chakraborty, \prl {\bf 65}, 108 (1990).

\bibitem{Que} W. Que, G. Kirczenow, and E. Castano,  \prb  {\bf  43}, 14079 (1991).







\bibitem{Ando} Y. Zheng and T. Ando, \prb {\bf  65}, 245420 (2002).

\bibitem{Jain} C. T\H{o}ke, P.\ E. Lammert, V.\ H. Crespi, and J.\ K. Jain,   \prb  {\bf   74}, 235417 (2006).

\bibitem{Lukose} V. Lukose, R. Shankar, and G. Baskaran, \prl {\bf 98}, 116802 (2007).

\bibitem{Gorkov} L.\ P. Gorkov and I.\ E. Dzyaloshinskii, JETP {\bf 26}, 449 (1967).

\bibitem{Lerner} I.\ V. Lerner and Yu.\ E. Lozovik, JETP {\bf 51}, 588 (1980).

\bibitem{Kallin}  C. Kallin and B.\ I. Halperin, \prb {\bf  30}, 5655 (1984);
\prb {\bf  31}, 3635 (1985).



\bibitem{Fertig} A. Iyengar, J. Wang, H.\ A. Fertig, and L. Brey,  \prb {\bf   75}, 125430 (2007).

\bibitem{gumbs} D. Huang and G. Gumbs, \prl {\bf B 46}, 4147 (1992).


\bibitem{Bottger} H. B\"{o}ttger and V.\ V. Bryksin, {\it Hopping Conduction in
Solids} (Akademie-Verlag, Berlin, 1985).

\bibitem{Thouless} D. J. Thouless, M. Kohmoto, M. P. Nightingale, and M. den Nijs, \prl {\bf 49}, 405 (1982).



\bibitem{LP}   E. M. Lifshitz and L. P. Pitaevskii, {\em  Course of Theoretical Physics\/},
 Vol. 9: Statistical Physics, Part II, (Moscow,
Nauka, 1978; Pergamon, New York, 1980).



\bibitem{Perov} V. Ya. Demikhovskii and A. A. Perov, \prb {\bf 75}, 205307 (2007).

\bibitem{gumbs2} G. Gumbs, D. Miessein, and D. Huang, \prb  {\bf 52}, 14755 (1995).

\bibitem{gumbs3} G. Gumbs and P. Fekete, \prb {\bf 56}, 3787 (1997).


\end{thebibliography}
\end{document}